\documentclass{PoS}
\usepackage[title]{appendix}
\usepackage{natbib}

\title{Broad-band Emission from Gamma-ray Binaries}

\ShortTitle{Broad-band Emission from Gamma-ray Binaries}

\author{\speaker{Josep M. Paredes} and {Pol Bordas} \\
              Departament de F\'{\i}sica Qu\`antica i Astrof\'{\i}sica, Institut de Ci\`encies del Cosmos, Universitat de Barcelona, IEEC-UB, Mart\'{\i} i Franqu\`es 1, 08028 Barcelona, Spain \\
        E-mail: \email{jmparedes@ub.edu}\\ 
        E-mail: \email{pbordas@ub.edu}}

\abstract{Gamma-ray binaries (GBs) have been object of intense studies in the last decade. From an observational perspective, GBs are phenomenologically similar to most X-ray binary systems in terms of their broad-band emission across the entire electromagnetic spectrum, being segregated from this source population by showing a maximum of their spectral energy distribution in the gamma-ray band, either at high-energies (HE: 100 MeV - 100 GeV) or very-high energies (VHE: above 100 GeV). From a theoretical perspective, the broad-band emission from GBs is a unique case in which particle acceleration and emission/absorption mechanisms can be tested against periodically changing conditions of their immediate surroundings. In this proceedings we examine some of the key observational results of the multi-wavelength emission from GBs. We discuss the correlated/contemporaneous emission observed in several of these systems, from radio to gamma-rays, by considering a single underlying particle-emitting population and the properties of the nearby photon, matter and magnetic ambient fields.}

\FullConference{Frontier Research in Astrophysics - III (FRAPWS2018)\\
		28 May - 2 June 2018\\
		Mondello (Palermo), Italy}

\begin{document}

% ---------------------------------------
\section{Gamma-ray binaries}
\label{section1}
% ---------------------------------------

In the last decade gamma-ray binaries (GBs) have arisen  as a new class of gamma-ray emitters at high-energies (HE; 100 MeV < $E$ < 100 GeV) and very-high-energies (VHE, $E $> 100 GeV). These systems are composed of a stellar-mass compact object orbiting a young, bright and massive star. GBs are segregated from their parent X-ray binariy population as they display non-thermal luminosities in the gamma-ray energy band at the level or higher than those observed at lower energy ranges, e.g. X-rays. As of today, seven systems have been firmly established as GBs: LS~5039 (\citealt{Aharonian2005_LS5039}), PSR~B1259-63 (\citealt{Aharonian2005_PSRB1259}), LS~I + 61~303 (\citealt{Albert2006}), HESS~J0632+057 (\citealt{Aharonian2007_HESSJ0632}), 1FGL~J1018.6-5856 (\citealt{LAT2012_1FGLJ1018}), LMC~P3 (\citealt{Corbet2016}), and PSR~J2032+4127 (\citealt{Lyne2015}). Table~1 summarises some key properties for these systems, including the orbital period, their extended morphology at radio wavelengths when present, their broad-band periodic or persistent emission, and the spectral type of the companion star and the nature of the compact object. For this latter item, only in the cases of PSR~B1259-63 and PSR~J2032+4127 the compact object has been firmly identified through the detection of pulsed emission when the neutron star is relatively far away from the free-free absorbing companion's stellar wind. The mechanism powering the non-thermal emission in the other cases is not yet definitely settled, with accretion-based models and pulsar-wind interaction scenarios still being discussed (see e.g. \citealt{Dubus2006, Paredes2006, Bosch-Ramon2006a, Romero2007, Torres2012, Jaron2016}.

% -----------------------------------------------------
\section{Multi-wavelength phenomenology of GBs}
\label{section2}
% -----------------------------------------------------

GBs are observationally characterised by their broad-band emission, with their spectral energy distribution peaking at gamma-ray energies. Almost all systems known so far display however distinct multi-wavelength features, evidencing the complex mechanisms at play behind the observed non-thermal emission. Below we briefly highlight some of the main multi-wavelength properties for the seven GBs known today, emphasising  the correlation of the emission at different bands in the cases it has been reported. For an in-depth review on some GBs discussed here ( but for LMC P3 and PSR J2032+057, as their gamma-ray detection has been reported only very recently) the reader is referred to \cite{Dubus2013}.

% ------------------------
\subsection{VHE gamma-rays}
\label{section2.1}
% ------------------------

At VHEs, all seven systems reported in Table~1 have been detected by the current third generation of Cherenkov telescopes. At this energy band, GBs appear as point-like sources (except for PSR J2032+057, for which the nature of the surrounding extended emission is still uncertain) with significant flux variability which in most cases is correlated with the system orbital motion. In PSR~B1259-63, the long period and the high eccentricity of the orbit explain the detection of the source only during orbital phases close to its periastron passage (\citealt{Aharonian2005_PSRB1259, Aharonian2009, HESS2013}, see also \citealt{Romoli2015}). Once detected, the TeV light-curve displays two broad maxima about 15 days before and after periastron, possibly associated to the two crossings of the neutron star through the companion's circumstellar disk. The GB system LS~5039 is detected all along the orbit (the orbital period is only $\sim$ 3.9 days), although it displays very different spectral properties when the system is found at superior or inferior conjunction (see e.g. \citealt{Aharonian2006}). LS~5039's orbital modulation has kept remarkably stable in the more than 10 years of monitoring the source at VHEs (see e.g. \citealp{Mariaud2015}). In the GB system LS~I + 61~303, the TeV emission is also linked to the orbital motion of the system. However, in contrast to LS~5039, the orbital phases in which the source is detected can change considerably depending on the epoch of observations (see e.g. \citealt{Albert2006, Acciari2008, Acciari2011}). The monitoring of HESS~J0632+057 at VHEs has unveiled the presence of two separate peaks in the source phase-folded flux profile, resembling those observed in X-rays (\citealt{Aliu2014}). This correlation seems to hold not only in the mid to long-term observations of the source, but also in short time-scales as well as in cycle-to-cycle (super-orbital)variability of the source \citep{Mukherjee2018}. Significant variability has been reported for 1FGL~J1018.6-5856 at VHEs as well, with hints for a periodical signal at a $\sim 3 \sigma$ level (\citealt{HESS2015}). In the case of LMC~P3, the only GB system detected in an external galaxy, the H.E.S.S. collaboration recently reported on the detection of variable emission from the source, which is in turn apparently phase-locked to the orbital period of the system. Periodicity could not be yet deduced from the TeV observations alone (\citealt{HESS2018}). For PSR~J2032+4127, the VERITAS and MAGIC collaboration have also recently reported the detection at VHEs (\citealt{VERITAS2018}) of this $\sim 40$--$50$~yr orbital-period system. The light-curve at VHEs displays a sharp flux drop about seven days after periastron passage, lasting just a couple of days. Such a dip is reminiscent to what observed in PSR~B1259-63, which had been interpreted as the result of strong adiabatic losses (see e.g. \citealt{Khangulyan2007, Kerschhaggl2011}), or attributed to photon-photon absorption effects (\citealt{Sushch2017}). The VHE spectrum of PSR~J2032+4127 shows in addition a statistically significant high-energy cutoff at about $1$~TeV when all available data is taken together. This cutoff is lowered to about $\sim 0.3$--0.6~TeV for an analysis of 2017 data prior to the source periastron passage. An analysis focused on the high-state emission right during the 2017 periastron passage provides instead no statistical preference for such a cutoff (\citealt{VERITAS2018}). So far, this is the second case after LS~5039 in which a spectral cutoff is observed in the VHE spectrum of a GB system.

% ------------------------
\subsection{HE gamma-rays}
\label{section2.2}
% ------------------------

At HE gamma-rays, all systems have been detected, and previous upper limits on HESS~J0632+057 by \cite{Malyshev2017} are now superseded by the recent results by \cite{Li2017}. The gamma-ray flux is found to be modulated with the orbital period of the system in most cases. For PSR~B1259-63, in addition, strong gamma-ray flares have been reported in the last three periastron passages of the source. Although similar, they also display cycle-to-cycle differences, e.g. the onset time of the flare, the presence of sub-peaks during the high-level emission state, and the total gamma-ray luminosity that the outburst can reach (see e.g. \citealt{Abdo2011, Caliandro2015, Tam2018}). In LS~5039 (\citealt{Abdo2009_LS5039}) and LS~I + 61~303 (\citealt{Abdo2009_LSI}), the HE spectrum follows a power-law with an exponential cutoff at a few GeVs, resembling that of HE gamma-ray emitting pulsars (HE emission is however not pulsed). A hint for a second gamma-ray component, arising at energies above $\sim$ few tens of GeV, has been reported for these two sources (\citealt{Hadasch2012}), as well as in HESS~J0632+057 (\citealt{Li2017}). The GB nature of 1FGL~J1018.6-5856 and LMC P3 were originally unveiled at HE gamma-rays, through a blind-search approach which firstly identified the source orbital period in this energy band (\citealt{LAT2012_1FGLJ1018}, \citealt{Corbet2016}). In the case of PSR~J2032+4127, although the source is detected at HE gamma-rays, no significant orbital variability has been reported during its last periastron passage in November 2017 (\citealt{Li2018}; note that these authors report on a drop of about $\sim 50$\% of the source HE gamma-ray flux right after periastron passage, albeit not statistically significant).

% ------------------------
\subsection{X-rays}
\label{section2.3}
% ------------------------

At X-ray energies, all seven GBs have been detected and in all cases the emission is modulated with the orbital period of the system. GBs display relatively hard spectra at X-rays, with no cutoffs, and no evidences for Fe-line or Compton reflection components. In PSR~B1259-63, the source displays a double peak asymmetric profile around the system periastron passage. Furthermore, $Chandra$ observations revealed extended X-ray emission near apastron (\citealt{Pavlov2011}). This emission has been interpreted as synchrotron emission from the pulsar wind when it is shocked by a clump of plasma ejected from PSR~B1259-63 at the time of the system periastron passage, in which the neutron star crosses the dense circumstellar disk of the companion star (\citealp{Pavlov2015}; see also \citealp{Barkov2016}). Interestingly, \cite{Bosch-Ramon2017} have recently conducted hydrodynamical simulations of HESS~J0632+057 showing that extended, moving X-ray emitting structures similar to those observed in PSR B1259-63 could also be expected in HESS~J0632+057. 

In the case of LS~5039, its X-ray flux is modulated on the orbital period, displaying again a very remarkable stability over years, similar to what is observed at VHEs, and further supporting the X-ray/TeV correlation of the source (see e.g. \citealt{Takahashi2009}). The X-ray spectrum of LS~5039, as well as in all other GBs is rather hard. This could be explained with radiation losses being dominated by IC in the KN regime, which would also lead to relatively softer TeV spectra (see e.g. \cite{Khangulyan2008}), as it is indeed observed in all GBs known so far. On the other hand, LS~5039's spectrum lacks any evidence for a spectral cutoff, with the spectrum extending up to 200 keV as detected with INTEGRAL (\citealt{Hoffmann2009}). 

The X-ray flux from  LS~I + 61~303 is also hard ($\Gamma \lesssim 2.0$) and without evidence for a cutoff up to $\sim 70$~keV (\citealt{Chernyakova2006}). LS~I's flux is also modulated with its orbital period of $\sim 26$d. In this case, however, super-orbital modulation produces a shift in the X-ray phase-folded peak. Short-term variability from LS~I + 61~303 may include the occurrence of two bursts observed with the Swift-BAT, resembling magnetar bursts (\citealt{Torres2012}). 

HESS~J0632+057 displays also a hard X-ray spectrum ($\Gamma \approx 1.6$). Its light-curve displays two peaks, a sharp one at orbital phases $\sim 0.3$ and a broader one at phases 0.6--0.9. A sharp dip in the X-ray light-curve is observed right after the first, main peak. Short-term variability has also been observed, with flux variations up to $\sim 40\%$ (\citealt{Hinton2009}). A double-peak X-ray profile has also been reported for the GB system 1FGL~J1018.6-5856 (\citealt{LAT2012_1FGLJ1018}), which displays otherwise also a hard spectrum  ($\Gamma \approx 1.7$) extending up to $\sim 40$~keV without evidence of a cutoff. 

The X-ray light-curve of LMC P3 is also modulated with the $\sim$10.3d orbital period of the system (\citealt{Corbet2016}). Its spectrum is hard, with $\Gamma \approx 1.3$ (note that in this case statistical errors are relatively high, $\Delta \Gamma = 0.3$, \citealt{Corbet2016}). The presence or absence of a cutoff has not been so far constrained by archival observations of the source. 

For the last GB system discovered, PSR~J2032+4127, \cite{Li2018} made use of Swift-XRT to monitor its recent periastron passage in November 2017. The X-ray light curve increased since mid-October 2017, displaying a sharp dip  during the periastron passage itself, followed by a post-periastron X-ray flare lasting for about 20 days, which could be a consequence of the Be stellar disk crossing by the pulsar (similar, e.g., to PSR~B1259-63).

% ------------------------
\subsection{Radio emission}
\label{section2.4}
% ------------------------

The seven GBs discovered to date display non-thermal radio emission at GHz frequencies of synchrotron origin, and show radio light-curves which are modulated with the orbital motion except for the case of LS~5039 (see e.g. \citealt{Marcote2015}, \citealt{Marcote2016}). The GBs that have been observed with VLBI exhibit in all cases a compact core as well as extended emission resolved at milliarcsecond scales (\citealt{Moldon2012}). In four GBs the bright star companion has a circumstellar decretion disk that produces emission lines, whereas in the other three cases the companion are bright O stars without the presence of a disk. 

The GBs in which the bright companion is a star featuring emission lines have revealed elongated, cometary tail structures in all cases except PSR J2032+057/MT91~213. This system is one of the two GBs for which the nature of the compact object is known (the other being PSR~B1259-63). In the case of PSR J2032+057, the system displays however a much longer orbital period, of about $\sim$40--50~yr. During its periastron passage in November 2017, it was expected that PSR J2032+4127 would develop an extended radio outflow forming a cometary tail behind the pulsar that would rotate as the neutron turned around the companion star, as seen  in other GBs (see e.g. \citealt{Moldon2011}). However, so far no VLBI map showing a radio structure has been made available. In the case of PSR~B1259$-$63, the strong interaction between the stellar wind and the pulsar wind near the periastron passage has been shown to produce an outflow extending several AU away from the system (\citealt{Moldon2011}). This is the first observational evidence that non-accreting pulsars orbiting massive stars can produce variable extended radio emission at AU scales. In the case of the GB system LS~I~+61~303, a peculiar characteristic first discovered in radio and later found at VHE is the presence of two distinct periodicities, one of 26.5~d and another of 1667 days. The first one is associated to the orbital period, whereas the second one has been suggested to depend of the Be star disc (see however \citealp{Massi2013}). In addition, early VLBI observations of this source showed an elongated structure that was interpreted as a jet-like feature (\citealt{Massi1993, Paredes1998}) but later AU-scale radio imaging with the VLBA revealed a cometary emission and a morphology varying dramatically at periastron, favouring a pulsar-wind scenario (\citealt{Dhawan2006}). Recent VLBA observations of the source mapped the elliptical trajectory of the radio emission, showing a good agreement with the previous maps (\citealt{Wu2018}). The GB system HESS~J0632+057 displays extended and variable radio emission at 50-100 AU scales with similar morphologies as those found in other GBs. However, a detailed VLBI monitoring along the orbit is still lacking.

In the other three GBs for which no circumstellar disk is present, only LS~5039, which has the shortest orbital period, has been deeply observed in radio at milliarcsecond resolutions. Its elongated radio structure discovered in the 2000's was interpreted under a microquasar scenario (\citealt{Paredes2000}). Later VLBI observations covering the orbit showed however periodic changes in the source radio morphology (\citealt{Moldon2012}), which were interpreted instead as a signature of a young non-accreting neutron star interacting with the wind of a massive O-type stellar companion (\citealt{Dubus2006}). In the case of1FGL~J1018.6$-$5856, the source displays compact radio emission on milliarcsecond scales. Extended emission on these angular scales has not been yet detected, mostly due to observational constraints, but more sensitive observations may lead to detect extended emission originating from the putative cometary tail as observed in other GBs (see e.g. \citealt{Marcote2018}). ATCA observations of CXOU~J053600.0$-$673507 show a modulation of the radio flux densities on the gamma-ray period and a point-like source (\citealt{Corbet2016}). Higher resolution observations are needed for the source to be resolved.

%1FGL~J1018.6$-$5856
%CXOU~J053600.0$-$673507

%___________________________________________________________________________

\begin{table}[t!]
{\small
\caption[]{Upated GBs table containing some of the main observational properties corresponding to the seven systems known so far. Sources are distinguished in terms of the spectral properties of the non-degenerate companion star present in the system, displaying or not emission lines associated to the presence of a circumstellar disk. A brief description of their phenomenological behaviour from radio to VHEs is also included (see footnote for details). Question marks are placed there where the nature of the compact object is still unknown, and where hints for periodicity or radio morphological structures need to be observationally confirmed.\\}
\label{detections} 
%\begin{center}
%\begin{tabular}{@{}l@{\hspace{0.07cm}}c@{\hspace{0.05cm}}cc@{\hspace{0.05cm}}c@{\hspace{0.05cm}}c@{\hspace{0.05cm}}c@{\hspace{0.05cm}}c@{\hspace{0.05cm}}c@{}}
\begin{tabular}{@{}l@{\hspace{0.07cm}}c@{\hspace{0.05cm}}cc@{\hspace{0.05cm}}c@{\hspace{0.25cm}}c@{\hspace{0.35cm}}c@{\hspace{0.05cm}}c@{\hspace{0.09cm}}c@{}}
%\begin{tabular}{lccccc}

\hline \noalign{\smallskip}
~~~~~~~~~Name  & System    & Orbital   &  Radio      &   \multicolumn{4}{c}{Multi-wavelength periodicity} \\    \cline{5-8}  
           &  type         & period   &  structure  &              &          &           &       \\ 
           &                  & (d)        &  (AU)         &  Radio  & X-ray & GeV   &  TeV  \\ 
\noalign{\smallskip} \hline \noalign{\smallskip}
\multicolumn{6}{c}{\bf Emission line star companion}\\
\noalign{\smallskip} \hline \noalign{\smallskip}

{\bf PSR~B1259$-$63} &  O9.5 Ve + NS  & 1237  & Cometary tail  & P & P & P & P \\
                                      &                          &           &    $\sim$ 120            &     &    &       &    \\

{\bf LS~I~+61~303} &  B0 Ve + ?  & 26.5  & Cometary tail ?  & P & P & P & P \\
                                 &                   &           &   $\sim$ 10 $-$ 700     &     &    &       &    \\

{\bf HESS~J0632+057} &  B0 Vpe + ?  & 320  & Elongated   & V & P & P? & P \\
                                      &                      &         &    $\sim$ 60          &     &    &       &    \\

{\bf PSR J2032+057/MT91~213} &  Be + NS  & 40-50 yr  & ? & D & D & D & D \\

\noalign{\smallskip} \hline \noalign{\smallskip}
\multicolumn{6}{c}{\bf Non-Emission line star companion}\\
\noalign{\smallskip} \hline \noalign{\smallskip}

%{\bf XTE~J1118+480}   & K7V$-$M0V    & 1.9  &  0.17  & 6.9$\pm$0.9 &  t &$-$& %$-$&$\le0.03$   \\
%$11^{\rm h}18^{\rm m}$10\rl 85 & +BH \\
%$+48^{\circ}02^{\prime}$12\pri 9          &       &    &    &   \\ 
{\bf LS~5039} &  O6,5 V((f)) + ?  & 3.9  & Cometary tail ?   & p & P & P & P \\    
                       &                           &        &    10 $-$ 1000         &                  &    &       &    \\
{\bf 1FGL~J1018.6$-$5856} &  O6,5 V((f)) + ?  & 16.5  &  ?  & P & P & P & P \\    
{\bf CXOU~J053600.0$-$673507} &  O5 III + NS?  & 10.3  &  ?  & P & P & P & D \\ 
    ~~~~~ (LMC P3)                         &                      &         &            &     &    &       &    \\

\hline
\end{tabular}
%\end{center}
{\small
Note: P: Periodic emission, p: Persistent emission, V: Variable emission, D: Detected }
}
\label{TABLE1}
\end{table}
%_________________________________________________________________________

% -----------------------------------------------
\section{Broad-band emission processes in GBs}
\label{section3}
% -----------------------------------------------

As summarised in Sect.~\ref{section2}, GBs have been observed to emit in the whole electromagnetic spectrum, from radio to VHE gamma-rays. Several non-thermal radiation processes can explain such emission. On the one hand, leptonic mechanisms including synchrotron, Inverse Compton (IC) and thermal/relativistic Bremmstrahlung can be very efficient processes in the GB framework. The first requires the presence of a magnetic field in the emitter zone, which in the case of GBs could come either from the magnetised plasma of the pulsar wind, or within the jet ejecta in accretion-based models. The magnetic field associated to the stellar wind or to the external medium around the binary system could also act as a ``seed'' field for synchrotron to operate, but this requires the emitting particles to escape the pulsar wind or jet regions, e.g. in the forward shock at the jet/wind tip when the ejecta is braked down by the swept-up external medium. Synchrotron emission from relativistic electrons produces a highly anisotropic radiation beamed towards the particle motion direction. In the ejecta reference frame, where the particle distribution is isotropic, the emission will also be approximately isotropic. On the observer frame, however, the bulk motion of the plasma may Doppler-boost this emission, which will make the source to appear more luminous/dimmer depending on whether the outflow is directed towards/away from the observer, with observed luminosities scaling as $L = L_{0} \delta^{n}$, where $L_{0}$ is the intrinsic luminosity,  $\delta$ the Doppler factor $\delta = \gamma^{-1} (1 - \beta cos\theta)^{-1}$, $\gamma = (1 - \beta^{2})^{-1/2}$ is the electron Lorentz factor,  and the index $n$ which accounts for the dependency on the geometry of the outflow and the spectral index of the particle population, typically $n \sim 2$ -- 3. 

IC emission can also be highly efficient in GBs, given the strong photon field provided by the companion star (a bright O or B spectral type star in all GBs known so far), or the circumstellar disk present in some sources (see Table~1). This mechanism is however highly anisotropic as well, as in this case it depends on the interaction angle between the electron and the incoming photon. For an emitter located in the vicinity of the compact object, as usually assumed in the case of GBs, this will produce a periodically modulated flux light-curve, reflecting the changing interaction conditions along the orbital motion, as far as the system orbital plane is oriented with some significant inclination by the observer. Note however that photon-photon absorption can also be very efficient at gamma-ray energies, producing a modulation as it also has a strong angular dependence between the gamma-ray IC photon and the soft photon coming from the IC target field. The net effect of this absorption is to redistribute the energy flux at the highest energies towards lower energy ranges. Great attention is needed therefore when attempting to interpret high-energy gamma-ray spectra from GBs, as this strongly anisotropic absorption needs to be considered on top of the intrinsically modulated spectrum produced through IC emission. In some instances, the absorption of VHE photons and the posterior redistribution to lower energies seems however limited by observations, since secondary particles created in the photon-photon absorption process would strongly radiate e.g. X-rays through synchrotron emission, which is not observed (see e.g. \citealt{Bosch-Ramon2008}). 
%The aforementioned Doppler-boosting effect for a relativistically moving plasma also applies here. 

The presence of a relatively dense matter field in the immediate surroundings of GBs can make Bremsstrahlung emission and hadronic interactions an efficient emission mechanism as well. However, thermal Bremsstrahlung has not been detected yet in GBs (see e.g. \citealt{Zabalza2011}). On the other hand, relativistic Bremsstrahlung is mainly considered when estimating the high-energy fluxes from large-scale outflows interacting with the surrounding medium, whereas at binary-system length-scales synchrotron and IC dominate. Finally, proton-proton interactions depend on the capability of GBs to produce relativistic protons. These protons can then interact with the companion stellar wind or the surrounding interstellar medium yielding neutral pions $\pi^{0}$ which in turn decay into gamma-rays and secondary leptons. Secondaries can then then re-emit via synchrotron and IC emission processes. Hadronic emission from GBs is not as widely considered as leptonic channels, partially because in the hypothesis that GBs are powered by pulsars, the wind of which is a plasma thought to be mainly composed by electrons and positrons. We will focus below on the synchrotron and IC leptonic emission processes only.

% -----------------------------------------------
\subsection{Constraining a single electron population model}
\label{section3.1}
% -----------------------------------------------

%_________________________________________________________________________
\begin{figure}[htb]
\begin{center}
\includegraphics[width=0.99\textwidth]{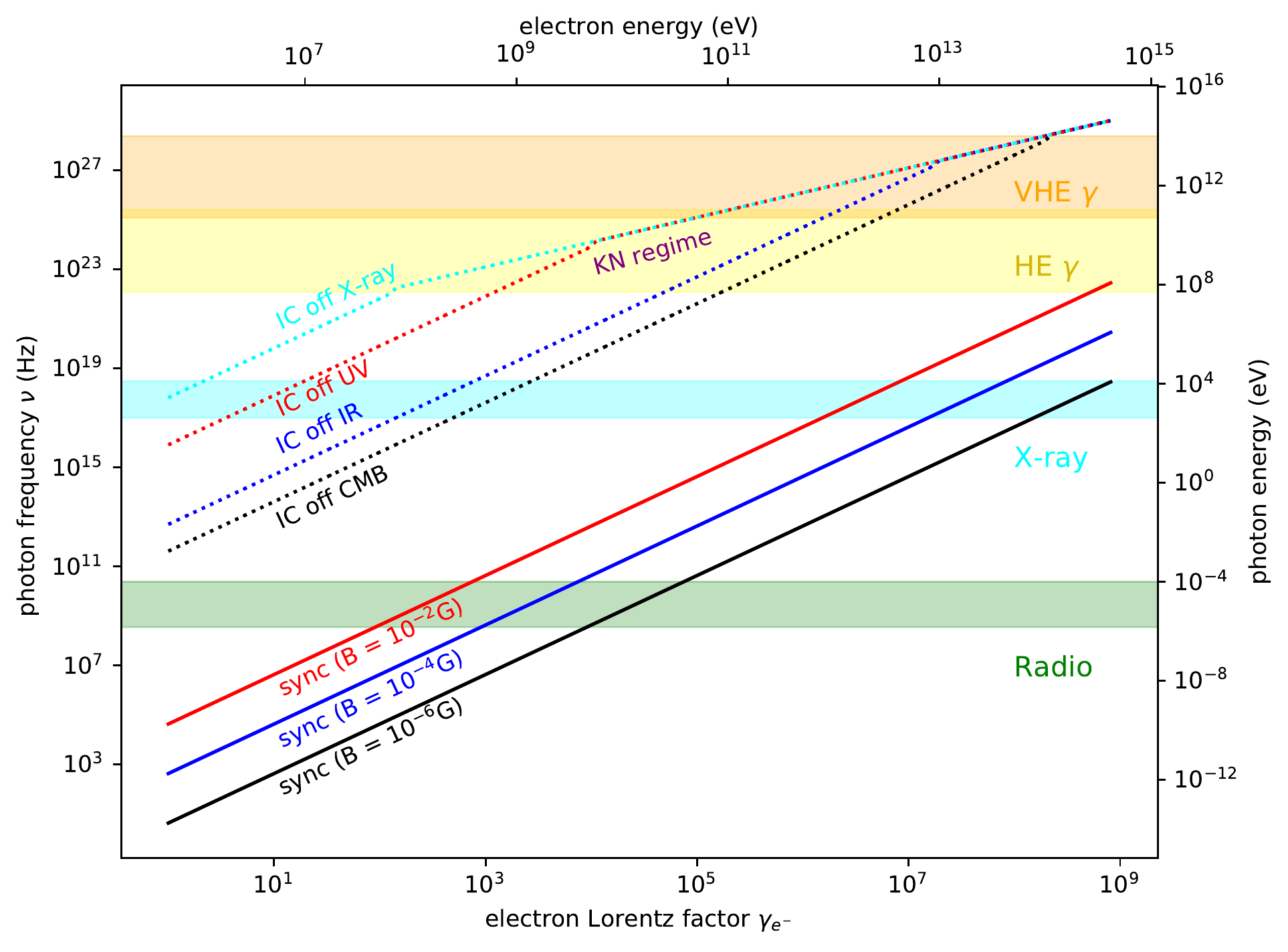}
\caption{Photon frequency (energy) produced through synchrotron and IC emission are plotted against the Lorentz factor (energy) of the emitting electrons for different values of the magnetic field and several seed photon fields found in GB systems and their surroundings. For IC, both TH and KN regimes are considered. Coloured bands indicate typical observational ranges employed in the study of GBs.}
\end{center}
\label{figure1}
\end{figure}
%_________________________________________________________________________

In the following we consider a simple one-zone model in which a single population may be responsible for the broad-band emission observed in GBs. Figure~1 displays the output photon frequency as a function of the electron Lorentz factor for synchrotron and IC emission processes. For the former, three values for the magnetic field are taken as a case-example, $B$ = 10 mG, 0.1 mG and 1$\mu$G. The outgoing photon frequency is taken as the critical frequency for synchrotron emission, $\nu_{C} = (\frac{3}{2}~$sin$\alpha) (E_{e}/m_{e}c^2)^2 (eB/2\pi m_{e}c) \propto E_{e}^{2}\,B_{\perp}$ (\citealt{Pacholczyk1970}). For IC we evaluate four different seed radiation fields given by the cosmic microwave background (CMB; $h\nu_{\rm CMB} \sim 10^{-15}$~erg) and the infrared emission component in the interstellar radiation field (ISRF; $h\nu_{\rm IR} \sim 1.2 \cdot 10^{-14}$~erg), as well as seed photons provided by the companion star and its circumstellar disk (or a transient accretion disk, see e.g. \citealp{Yi2017}), when present, at UV and X-ray frequencies, respectively ($h\nu_{\rm UV} \approx 10^{-11}$~erg and $h\nu_{\rm X} \approx 3.2 \cdot 10^{-9}$~erg). For IC we also consider the Thomson (TH) and Klein-Nishina (KN) regimes, for which a transition is taken from the former to the latter when $4\,\epsilon_{0}\,\gamma \gtrsim 1$, where $\epsilon_{0}$ is the seed photon energy in $m_{e}c^2$ units and $\gamma$ is the electron Lorentz factor (\citealt{Moderski2005}). Coloured bands in Figure~1 represent the energy-bands traditionally employed in the study of non-thermal emission from GBs, from radio to VHE gamma-rays. 
\vspace{0.5cm}

High-frequency radio emission from GBs is attained through synchrotron emission from electrons with Lorentz factors in the range $\gamma \in [10^3$--$10^5]$ for the choice of $B_{\perp}$ used here. For these magnetic field values, higher-energy electrons with $\gamma \in [10^7$--$10^9]$ will instead radiate into the X-ray band. For very energetic particles, with $\gamma \gtrsim 10^9$, a strong magnetic field could make GBs to be detected at HE gamma-rays (note however that electron-synchrotron radiation features a so-called "universal" cutoff at about 160 MeVs when Doppler-boosting effects are not accounted for, see e.g. \citealt{Aharonian2000}). 

Low-energy electrons, with $\gamma \in [10$--$10^3]$, can also produce X-ray emission through IC in the TH regime, whereas IC by higher-energy electrons, with $\gamma \in [10^3$--$10^6]$, will produce HE gamma-rays. For the X-ray seed photon field considered here, however, KN effects will become already substantial at electron Lorentz factors $\sim$ few $\times 10^2$. For less energetic seed photon fields, this transition occurs at increasingly higher electron energies (e.g. for the UV radiation field $\gamma \gtrsim 10^3$ are required). Scattering off the IR and CMB photon fields will instead produce HE gamma-ray emission while still well in the TH regime. Very high-energy electrons, with  $\gamma \gtrsim 10^7$, will produce VHE gamma-rays. Here again, IC will be deep into the KN regime for the high-energy seed photon fields, whereas the TH-KN transition will take place somewhere in the VHE band for IR seed photons, and at the edge of the VHE observational window for the CMB, for output gamma-rays above few $\times 10$~TeV. 

For the X-ray seed photon field employed in Figure~1, the KN effect makes the range of energies of electrons that emit gamma-rays to be relatively wide, spanning from $10^3$ to $\lesssim 10^6$ for HEs and from $\sim 10^6$ to $10^8$ for VHEs. If separate electron populations are responsible for the gamma-ray emission in GBs produced by the up-scattering of such X-ray photon field at HE and VHEs, these electron populations should feature a cutoff at about $\Gamma_{e}\sim 10^{5}$ (a high-energy cutoff for the less energetic population responsible for the HEs, and a low-energy cutoff in the distribution of electrons responsible for the VHE emission). It is also worth noting that for a relativistic electron population injected following a power-law $N(E) \propto E^{-p}$ with a spectral index $p \sim 2$, the spectral energy distribution is flat, $E^{2} \times N(E) = $ constant, and the total accumulated energy in particles is proportional to the width of the energy interval considered, $\Delta E$. In this regard, the wide range of energies spanned by the electrons that produce HE or VHE gamma-rays through IC on the X-ray photon field makes this observational energy window one of the most efficient ones to cover and quantify the non-thermal particle kinetic luminosity of the system. 

The presence of such strong X-ray field within or surrounding a given GB system may not be however necessarily common. In an exercise aiming to constrain the kinetic energy output of the GB system LS~5039, \cite{Zabalza2011} used the expected thermal X-rays produced in a colliding wind zone scenario for this source to derive an upper limit to LS~5039's spin down luminosity. Signatures of thermal X-ray emission are not observed in general for GBs. X-ray emission could be otherwise associated to accreting processes. Here again the corresponding radiative signatures from GBs are lacking, a fact that has played a role in the interpretation of GB systems as being powered by pulsar winds (\citealt{Martocchia2005, Dubus2006}). Non-thermal X-rays could be nevertheless produced in this latter scenario by electrons accelerated at the wind-wind shock interface (via synchrotron) or in the close encounter of the compact object and a companion star featuring a dense circumstellar disk, in particular for systems displaying high eccentricity orbits. Some GBs, like the system PSR~B1259-63, display contemporaneous radio, X-ray and TeV emission close to periastron, with similar double-peaked flux profiles (see e.g.\citealp{Chernyakova2014, Kerschhaggl2011}; see also Sect.~\ref{section2}). This emission could be originated from the same colliding-wind shock-accelerated electron population (\citealt{Maraschi1981, Tavani1997, Kirk1999, Khangulyan2007, Bosch-Ramon2011}). PSR~B1259-63 displays however also strong gamma-ray flares a couple of weeks after periastron passage, which so far have been detected at HEs only (see Sect.~\ref{section2}). The fluxes at X-rays and VHEs do not display any sudden increase similar to such flares. Still, the contemporaneous X-ray and VHE flux levels are relatively high (see e.g. \citealt{Chernyakova2015}). At VHEs in particular, they can be at the level of those found right at the time that the neutron star crosses the dense equatorial disk (\citealt{Romoli2015}). Other scenarios, in which HE gamma-rays are Doppler-boosted synchrotron photons (see e.g. \citealt{Kong2012}) cannot be discarded either. In this case, however, the emission at other wavelengths should also be Doppler-boosted. 

Synchrotron X-rays and IC VHE gamma-rays are often assumed to be produced by a common particle distribution. Figure~1 shows that, for electron Lorentz factors of $\sim 10^7$--$10^9$, synchrotron X-rays require magnetic fields of at very least few $\mu$G, but most probably few mG or more. Gamma-rays, on the other hand, would be produced by the same parent population through IC on the strong photon field from the companion star present in all seven GBs known so far. A contribution from external fields or the self-produced X-ray emission (see above) could also be at play in a scenario in which a single distribution is responsible for the X-ray and TeV emission. 

HE gamma-rays detected from GBs seem to require instead separate populations in a number of cases. This conclusion can be derived e.g. upon inspection of the spectral energy distribution obtained with the $Fermi$-LAT for the systems LS~I + 61~303 and LS~5039 (see e.g. \citealt{Hadasch2012}; see also \citealt{delPalacio2015} for a study focused on LS~5039). Hints for independent emitting particle populations being responsible for the HE and VHE emission have also been posed in the recent LAT results on HESS~J0632+057 by \cite{Li2017}, the gamma-ray SED of which seems difficult to explain with a single leptonic distribution. Further examples of the non-correlation of HE emission with other windows are provided e.g. by the aforementioned strong gamma-ray flare observed in PSR~B1259-63, with the HE flux increasing by a factor $\gtrsim 9$ in a few days that has no similar counterpart at any other wavelength.

%NOTE: synchrotron self-compton should balance the efficiency of the two channels. If synch radiation losses dominate and emit strongly, the seed photon becomes more  and more ense, at some point so dense that IC becomes more effective. On the contrary, IC cannot stay much longer being so effective since it depends on synchrotron. 

%Note that the synchrotron burn-off limit at about 160 MeV is not represented in Fig. 1. 

% ---------------------------------
\section{Conclusions}
\label{section4}
% ---------------------------------

GBs display non-thermal emission spanning a broad energy range, from radio to gamma-ray energies. This emission is modulated with the orbital period except for a few non-confirmed cases (e.g. LS~5039 at radio wave-lengths and PSR~J2032+4127 at HE gamma-rays). Light-curves can also display distinct features on a source-by-source basis, including among others super-orbital modulation signatures, asymmetric double-peak profiles, sharp dips, and strong flares. GB spectra are usually hard at radio/X-rays, softening somewhat at gamma-ray energies. In this latter regime, different particle populations and/or emitting regions seem to be required to explain the HE-VHE emission. Despite that only a few GBs are known so far, an accurate modelling of their broad-band emission is complex and requires a detailed treatment which may differ from source to source, akin to their distinct observational properties.  

~\\

{\bf Acknowledgements}
This work was supported by the Agencia Estatal de Investigaci\'on grant AYA2016-76012-C3-1-P from the Spanish Ministerio de Econom\'{\i}a y Competitividad (MINECO); by grant MDM-2014-0369 of the ICCUB (Unidad de Excelencia 'Mar\'{\i}a de Maeztu'); and by the Catalan DEC grant 2017 SGR 643, as well as FEDER funds. Pol Bordas acknowledges the finantial support by the Beatriu de Pin\'{o}s Research Program from the Agency for Management of University and Research Grants (AGAUR) of the Generalitat de Catalunya. 

%\bibliography{bibliography}
\bibliographystyle{aa} {\footnotesize\bibliography{bibliography}}

\clearpage
\begin{appendices}
\section{Questions and Answers}
\begin{itemize}
\item{\bf JIM BEALL QUESTION:} Do you have evidence of variability in a jet that could be generated by the neutron star and deccretion disk interaction? 

\noindent ANSWER: in the context of GBs, extended synchrotron radio emission has been recently interpreted as the cometary tail produced the shocked pulsar wind in its motion around the companion star. Such interpretation benefits from the observational fact that this extended emission seems to change its position angle as a function of the system orbital phase. Yet, accretion-powered jet models cannot be ruled out, but they would require a strong precession in order to reproduce this orbitally-modulated position-angle change. Jet models for the system LS~I~+61~303 keep being proposed nowadays, e.g. by M. Massi and collaborators. This source in particular may display a deccretion disk. Jet precession may facilitate the jet/disk interaction formulated in this question. Regular outbursts following enhanced particle acceleration in the jet/disk shock would be expected in such a case, although their prediction would strongly depend on the system geometry and the actual properties of both jet and disk.  
\item{\bf WOLFGANG KUNDT:}	What are your objections to my 1989 paper with Daniel Fisher: "Is there a blackhole among the black-hole candidates?" in Journal of Astrophysics and Astronomy, 10, 119 (1989).

\noindent ANSWER: for the topic on the broad-band emission from gamma-ray binaries discussed in this contribution we have no objections to the black-hole or neutron star nature for the compact object. In this regard, it is worth noting that at least in two cases the detection of radio pulsations precludes the presence of a black-hole (in PSR~B1259-63 and PSR~J2032+4127, both discussed in these proceedings).
\end{itemize}
\end{appendices}

\end{document}